\documentclass{article}

\usepackage{arxiv}

\usepackage[utf8]{inputenc} 
\usepackage[T1]{fontenc}    
\usepackage{hyperref}       
\usepackage{url}            
\usepackage{booktabs}       
\usepackage{amsfonts}       
\usepackage{nicefrac}       
\usepackage{microtype}      
\usepackage{lipsum}		
\usepackage{graphicx}
\usepackage{natbib}
\usepackage{doi}

\title{Navigating the Knowledge Sea: Planet-scale answer retrieval using LLMs}

\author{ \href{https://orcid.org/0000-0001-5431-6367}{\includegraphics[scale=0.06]{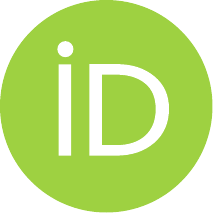}\hspace{1mm}Dipankar Sarkar} \\
  Terraprompt AI \\
  \texttt{me@dipankar.name} \\
}

\renewcommand{\shorttitle}{Navigating the Knowledge Sea: Planet-scale answer retrieval using LLMs}

\hypersetup{
pdftitle={Navigating the Knowledge Sea: Planet-scale answer retrieval using LLMs},
pdfsubject={cs.IR,cs.LG},
pdfauthor={Dipankar Sarkar},
pdfkeywords={Information Retrieval,Large Language Models,Search},
}

\begin{document}
\maketitle

\begin{abstract}
Information retrieval is a rapidly evolving field of information retrieval, which is characterized by a continuous refinement of techniques and technologies, from basic hyperlink-based navigation to sophisticated algorithm-driven search engines. This paper aims to provide a comprehensive overview of the evolution of Information Retrieval Technology, with a particular focus on the role of Large Language Models (LLMs) in bridging the gap between traditional search methods and the emerging paradigm of answer retrieval. The integration of LLMs in the realms of response retrieval and indexing signifies a paradigm shift in how users interact with information systems. This paradigm shift is driven by the integration of large language models (LLMs) like GPT-4, which are capable of understanding and generating human-like text, thus enabling them to provide more direct and contextually relevant answers to user queries. Through this exploration, we seek to illuminate the technological milestones that have shaped this journey and the potential future directions in this rapidly changing field.
\end{abstract}


\section{Introduction}
The landscape of information retrieval has undergone transformative changes over the past few decades, driven by relentless advancements in computing technology and innovative algorithmic approaches. This evolution, marked by a gradual yet profound shift from basic information organization to sophisticated search and retrieval mechanisms, underscores a persistent quest to satisfy the ever-growing human thirst for knowledge and information efficiency.

In the early stages of the World Wide Web, information organization was predominantly manual, characterized by web pages serving as rudimentary repositories of data \citep{berners1994world}. The subsequent advent of directories, exemplified by Yahoo! Directory, represented a pivotal moment in information categorization based on popularity and relevance \citep{hock2005yahoo}. These directories laid the groundwork for more dynamic information retrieval methods.

The development of initial search engines, like AltaVista and Lycos, represented a substantial advancement in the availability of information. These platforms utilized improved computational capabilities to index and retrieve web content in a more efficient manner compared to previous method. Nevertheless, it was the introduction of Google's search engine, incorporating the innovative PageRank algorithm, that fundamentally transformed the process of retrieving information. Google's approach of distributed computing facilitated thorough web crawling, resulting in nearly instantaneous and extensively comprehensive access to information \citep{brin1998anatomy}.

Despite these advancements, users were still required to engage in considerable manual effort to sift through links and extract pertinent information. This challenge set the stage for the integration of Large Language Models (LLMs) in information retrieval systems. The development of models such as OpenAI's GPT (Generative Pretrained Transformer) series, culminating in GPT-4, represents the latest frontier in this evolutionary journey \citep{brown2020language}. These LLMs, by virtue of their deep learning architectures, are capable of understanding and generating human-like text, thus enabling them to provide more direct and contextually relevant answers to user queries.

Modern services like Perplexity.ai \citep{PerplexityAI2023} and Bing AI Search \citep{BingAI2023}, powered by these advanced LLMs, illustrate a significant paradigm shift. They exemplify how the integration of LLMs over existing search indexes can generate precise answers by analyzing link contents, thereby reducing user effort and enhancing the overall search experience \citep{fostikov2023first}.

However, this evolution is not without its challenges. The prevalence of spam links and the need for accurate context interpretation necessitate a rethinking of web crawling and indexing strategies. The potential of LLMs in directing crawls and selectively filtering links based on user intent is an area of ongoing research and development \citep{bethany2024large}.

This paper aims to provide a comprehensive overview of the evolution of information retrieval technology, with a particular focus on the role of Large Language Models in bridging the gap between traditional search methods and the emerging paradigm of answer retrieval. Through this exploration, we seek to illuminate the technological milestones that have shaped this journey and the potential future directions in this rapidly evolving field.

\section{Historical Perspective of Information Retrieval}

\begin{figure}[h]
    \centering
    \includegraphics[width=0.5\linewidth]{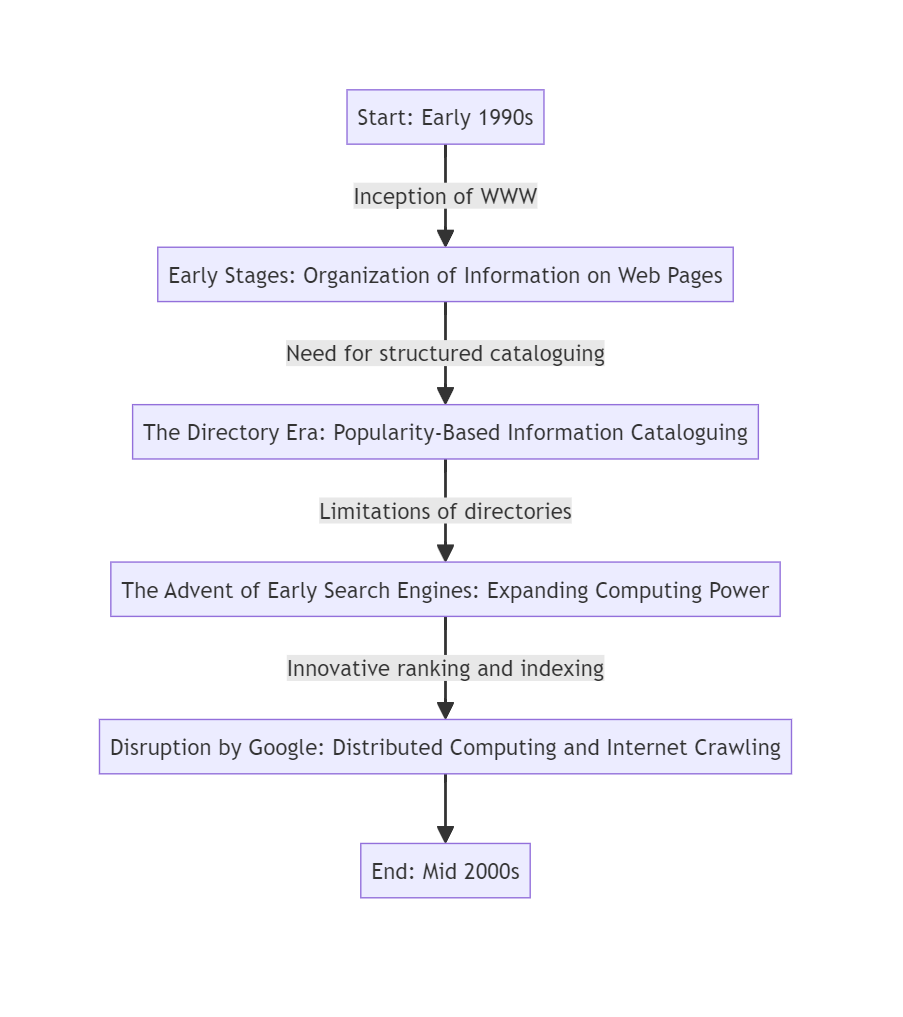}
    \caption{Evolution of internet information retrieval}
    \label{fig:information-retrieval}
\end{figure}

The journey of information retrieval systems is a testament to the remarkable evolution of digital technology and its profound impact on the accessibility of information. This historical trajectory, from rudimentary web page organization to the advent of sophisticated search engines, mirrors the exponential growth in both the volume of digital information and the complexity of user needs.

The development of information retrieval throughout history has involved a consistent improvement of methods and technologies. This progress has ranged from simple hyperlink-based navigation to advanced search engines powered by algorithms. Each phase of this evolution has been motivated by the increasing complexity and abundance of information on the internet, with the ultimate aim of enhancing accessibility and usability for users.

\paragraph{Early Stages} The inception of information retrieval can be traced back to the early 1990s with the advent of the World Wide Web, conceptualized by Tim Berners-Lee. In its nascent stages, the web was a collection of static pages linked by hyperlinks, a digital space of information with minimal organizational structure \citep{berners1994world}. This era was characterized by the manual curation of web content, often limited in scope and accessibility.

\paragraph{The Directory Era} As the web expanded, the need for structured cataloguing became apparent. This led to the creation of directories, such as Yahoo! Directory, which organized websites into categories, simplifying the user's search process. These directories relied on human editors to evaluate and categorize web content, representing an early attempt to impose order on the burgeoning web.

\paragraph{The Advent of Early Search Engines} The limitations of manually curated directories soon became evident with the exponential growth of the web. This challenge was met by the development of the first search engines, like AltaVista and Lycos, which automated the process of information retrieval. These search engines utilized rudimentary algorithms to index web pages and provided a basic search functionality, marking a significant advancement in information retrieval \citep{bar2006methods}.

\paragraph{Disruption by Google} A watershed moment in the history of information retrieval was the launch of Google in 1998. Founders Larry Page and Sergey Brin introduced the PageRank algorithm, which used the link structure of the web as an indicator of a page's importance, revolutionizing the way search results were ranked \citep{brin1998anatomy}. Google's distributed computing approach enabled the comprehensive indexing of the web, significantly improving the speed and relevance of search results, setting a new standard for information retrieval.

\section{User Interaction with Information Retrieval Systems}

\begin{figure}[h]
    \centering
    \includegraphics[width=0.4\linewidth]{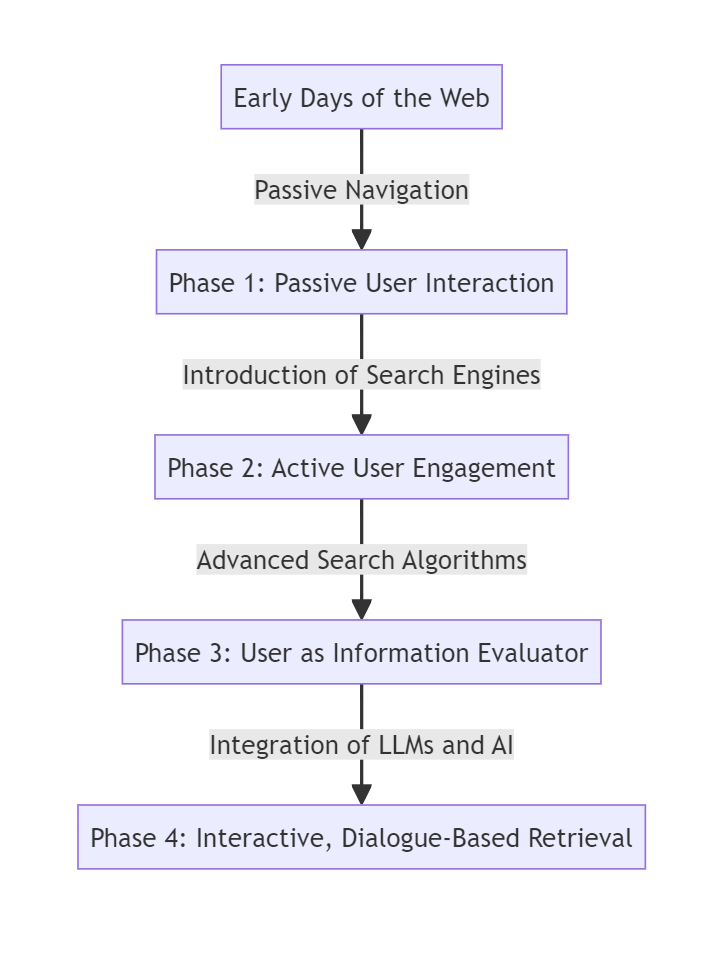}
    \caption{Evolution of user interaction}
    \label{fig:user-interaction}
\end{figure}

The evolution of information retrieval systems is not only a story of technological advancement but also a narrative of changing user interactions and behaviors. This section delves into the transformative journey of how users have engaged with these systems, from the early days of the web to the present.

The development of user interaction with information retrieval systems demonstrates an ongoing adjustment to technological progress. This transition from passive browsing to active involvement and, eventually, to interactive conversation highlights a broader change in the users' role in the information retrieval process, moving from simply seeking information to becoming critical assessors and participants in discussions.

\paragraph{Evolution of User Engagement} In the initial phase of the World Wide Web, user interaction with information was predominantly passive. Early web users navigated through a limited number of hyperlinked documents, often lacking sophisticated tools for efficient information retrieval \citep{nielsen1999user}. This era was marked by a reliance on the hierarchical structure of directories, where the user's role was largely confined to navigating through categorically arranged links \citep{koster1994aliweb}.

\paragraph{Limitations of Traditional Search Engines} The advent of search engines transformed user interaction by enabling more active engagement. Users could input queries and receive a list of relevant web pages. However, this process often required sifting through numerous results to find pertinent information, a task that could be both time-consuming and overwhelming \citep{marchionini2006exploratory}. The reliance on keyword-based search queries also posed challenges, as it required users to accurately anticipate and use the terms used in the content they were seeking.

\paragraph{Personalisation of search} The introduction of more sophisticated search algorithms and personalized search results further refined user interactions. Google's PageRank algorithm, for instance, not only improved the relevance of search results but also subtly shifted the user's role from an information seeker to an information evaluator, discerning the credibility and relevance of the presented links \citep{brin1998anatomy}.

\paragraph{Search dialog systems} With the integration of Large Language Models (LLMs) in search systems, user interaction has undergone another significant transformation. Modern systems, equipped with AI like GPT-4, are not only capable of understanding complex queries but also of providing direct, conversational responses. This advancement represents a shift towards a more interactive, dialogue-based form of information retrieval, where the system understands the context and intent behind user queries, thereby reducing the cognitive load on the user.

\section{Large Language Models in Information Retrieval}

The incorporation of large language models (LLMs) like GPT-4 into information retrieval represents a significant progress in the field. Retrieval Augmented Generation (RAG) is a crucial aspect of this integration, as it combines the capabilities of LLMs with dynamic information retrieval to improve the precision and pertinence of the generated responses.

The evolution from GPT to GPT-4 exemplifies major strides in natural language processing (NLP). These LLMs are built on sophisticated deep learning architectures, enabling them to comprehend and generate human-like text, significantly enriching user interaction with information retrieval systems \citep{brown2020language}.

RAG represents a novel approach in information retrieval, wherein the LLM not only generates responses but also retrieves and incorporates relevant external information in real-time. This mechanism allows the model to dynamically pull in data from a wide range of sources, ensuring that the answers provided are both current and contextually rich \citep{lewis2020retrieval}. RAG essentially combines the generative capabilities of models like GPT-4 with the information retrieval prowess of traditional search engines, leading to more accurate, informative, and up-to-date responses.

In practice, services such as Perplexity.ai \citep{PerplexityAI2023} and Bing AI Search \citep{BingAI2023} utilize LLMs enhanced with RAG to deliver a more sophisticated search experience. By leveraging RAG, these platforms are able to understand complex queries and generate responses that are not only contextually aware but also infused with the most relevant and recent information available across the web. This approach significantly surpasses traditional search methodologies, providing users with a comprehensive, concise, and highly informative answer \citep{fostikov2023first}.

Case studies focusing on the implementation of RAG in search technologies reveal a substantial improvement in user satisfaction. Users benefit from responses that are not only semantically aligned with their queries but also enriched with the latest information, offering a depth of understanding previously unattainable in standard search engines \citep{gao2023retrieval}.

The combination of LLMs, especially when combined with Retrieval Augmented Generation, signifies a significant transformation in the field of information retrieval. This advancement brings together generative language models and conventional search methods, offering users an unparalleled level of precision, pertinence, and timeliness in their search outcomes.

\section{Technology Shift from Link Retrieval to Answer Retrieval}

The field of information retrieval has been profoundly influenced by technological advances, particularly in the realm of AI and machine learning. This section explores the significant technological developments that have shaped the modern landscape of answer retrieval, focusing on how these innovations have refined the process of information discovery and extraction.

Historically, information retrieval systems primarily focused on link retrieval, presenting users with a list of web pages relevant to their queries. This paradigm began to shift with the advent of more sophisticated algorithms and AI technologies. The emergence of semantic search technologies, which understand the meaning and context behind user queries, marked a significant departure from traditional keyword-based searches \citep{schutze2008introduction}. This shift towards semantic understanding laid the groundwork for systems that could provide direct answers instead of mere links.

As the internet continued to grow, the prevalence of spam links and irrelevant data emerged as significant challenges in information retrieval. The development of advanced filtering algorithms and machine learning techniques has been crucial in addressing these challenges. For instance, Google's introduction of the Panda and Penguin updates to its algorithm significantly improved the quality of search results by penalizing low-quality and spammy content \citep{patil2021comparative}. These technological advancements have been instrumental in enhancing the relevance and accuracy of answer retrieval.

The integration of Large Language Models (LLMs) like GPT-4 into search systems represents the latest frontier in the evolution of answer retrieval. LLMs offer unprecedented capabilities in understanding natural language, context, and user intent. This allows for more effective crawling strategies, where crawlers can be directed to focus on content that is more likely to be relevant to future queries. Additionally, LLMs can be employed to perform more nuanced filtering of search results, ensuring that the answers provided to users are not only relevant but also of high quality and reliability \citep{brown2020language}.

In conclusion, the continuous technological advances in the field of information retrieval have significantly enhanced the process of answer retrieval. From the early stages of link-centric searches to the current AI-driven, answer-focused systems, these developments have continuously strived to make information more accessible, relevant, and reliable for users.

\section{Planet-scale Answer Retrieval with Large Language Models}

\begin{figure}[h]
    \centering
    \includegraphics[width=1\linewidth]{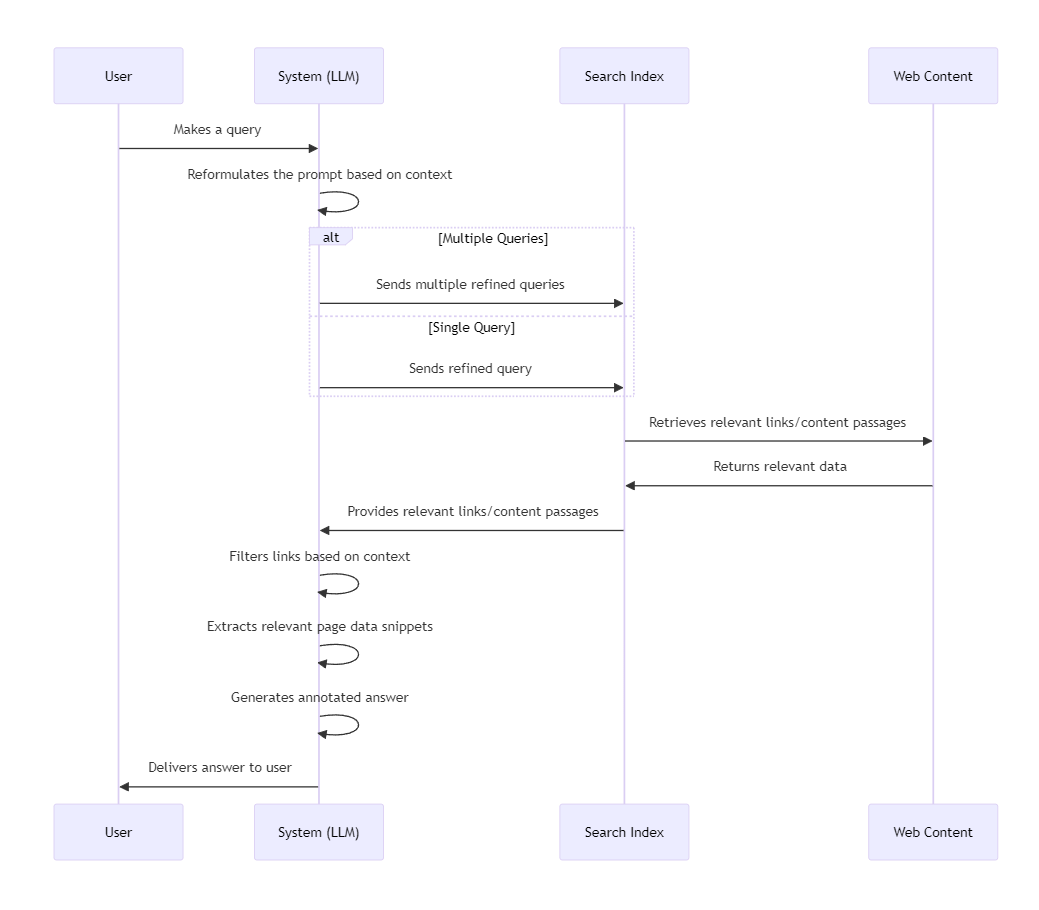}
    \caption{Steps for planet-scale answer retrieval}
    \label{fig:planet-answer-retrieval}
\end{figure}

Incorporation of large language models (LLMs), such as GPT-4, into the process of retrieving answers signifies a notable progress in the domain of information retrieval. This section explores the intricacies of this process, elucidating how LLMs have transformed the manner in which information is accessed and delivered to users. The answer retrieval process with LLMs can be delineated into six distinct but interconnected steps \ref{fig:planet-answer-retrieval}:

\begin{itemize}
    \item \textbf{User make a query}  The process begins with the user posing a query. LLMs, with their advanced natural language understanding, interpret these prompts not just by the keywords but by grasping the contextual and semantic nuances of the query .
    \item \textbf{System Reformulating the Prompt Based on Context} The LLM then reformulates the query to better capture the user's intent for the index search, it might result in multiple queries for the index. This step involves a sophisticated understanding of the query's context, possible ambiguities, and the specific information need .
    \item \textbf{System Calling the Index} Once the query is refined, the system accesses an extensive search index. We already see the usage of planet-scale indexes like google, bing. Alternatively, the system can crawl the internet and build its own index. This can return relevant links, which need to be scraped or even relevant content passages. 
    \item \textbf{System Filtering Links Based on Context} The LLM filters the retrieved information, prioritizing results based on relevance to the reformulated query. This filtering is crucial in eliminating spam and irrelevant links, ensuring that only the most pertinent information is considered .
    \item \textbf{System Extracting Relevant Page Data Snippets} The LLM then extracts snippets of information from the filtered results. This involves parsing through the content to identify and extract sections that directly address the user's query.
    \item \textbf{System Generating Annotated Answer}: Finally, the system synthesizes the extracted information into a coherent, concise, and contextually relevant answer. This step often includes the generation of annotations or references, providing the user with insights into the source and reliability of the information.
\end{itemize}

The ability of LLMs to understand and incorporate context in query reformulation is a critical component of the answer retrieval process. By considering the user's search history, the nature of the query, and even the broader societal and linguistic context, LLMs provide a level of query refinement that is significantly more advanced than traditional keyword-based systems \citep{schutze2008introduction}.

LLMs have brought forth novel techniques for extracting and annotating answers from extensive datasets. In contrast to previous systems that heavily relied on keyword matching, LLMs employ deep learning to comprehend the core of the content, allowing them to extract information that is more precise and pertinent. Additionally, the annotations generated by LLMs provide users with supplementary context, thereby enhancing transparency and fostering trust in the retrieved information.

The utilization of Large Language Models in the answer retrieval process signifies a revolutionary method for obtaining information. By harnessing the advanced functionalities of LLMs, this process not only improves the effectiveness and precision of information retrieval but also enhances the overall user experience by delivering contextually comprehensive and well-informed answers.

\section{Large Language Models based indexing}

The integration of large language models (LLMs) such as GPT-4 into the indexing process of information retrieval systems marks a new era in how digital content is organized, processed, and retrieved. This section examines the transformative impact of LLMs on indexing methodologies and the implications for future information retrieval systems.

The advent of LLMs has significantly influenced the techniques employed in web page crawling. Traditional web crawlers relied on basic algorithms to navigate and index web content. However, with LLMs, the process becomes more nuanced and intelligent. LLMs can understand the context and relevance of web content more deeply, enabling crawlers to prioritize and categorize pages more effectively based on their potential usefulness in answering user queries \citep{brown2020language}.

The PageRank algorithm, a foundational technology in Google's indexing process, has historically been pivotal in determining the importance of web pages based on their link structure. In the age of LLMs, this approach can be augmented with more sophisticated scoring methods. LLMs can analyze the quality of content, the relevance of the information to potential queries, and even the credibility of sources. This enhanced scoring capability allows for a more refined and relevant ranking of web pages in search results \citep{brin1998anatomy}.

LLMs have also introduced new possibilities in selective filtering and index enhancement. By understanding the nuances of language and user intent, LLMs can filter out irrelevant, low-quality, or spammy content more effectively during the indexing process. This selective filtering ensures that the indexed content is not only relevant but also of high quality, thereby improving the overall efficiency and effectiveness of the search process.

In addition, LLMs will contribute to the enhancement of the index by enabling the inclusion of semantic and contextual metadata in the indexing process. This metadata enriches the index, allowing for more sophisticated and nuanced retrieval that aligns closely with the user's search intent and context.

The utilization of LLMs in the indexing procedure shall signify a notable progress in the domain of information retrieval. Through the integration of deep learning and natural language understanding into conventional indexing methods, LLMs have the capability to transform information organization and retrieval in the digital era, resulting in increased relevance, efficiency, and user-friendliness.

\section{Conclusion}

This paper explores the development of information retrieval, tracing its progression from basic web directories to the advanced use of Large Language Models (LLMs) like GPT-4 in modern search systems. The evolution of information retrieval mirrors the broader trajectory of technological advancement, emphasizing a constant drive towards more efficient, accurate, and user-focused approaches to accessing information. 

The incorporation of LLMs in information retrieval, particularly in answer retrieval and indexing, represents a significant shift in how users engage with information systems. LLMs have transformed search from a process of sifting through links to actively participating in a conversation and context-aware exchange. This shift not only improves the efficiency and relevance of search results but also fundamentally changes the user's role, elevating them from information seekers to informed participants in a dialogue-driven retrieval process. 

Additionally, the use of LLMs in indexing wil lredefine the organization and accessibility of digital content. By enabling more intelligent and selective indexing methods, LLMs ensure that the vast amount of online information is curated in a relevant and reliable manner. This advancement is crucial in an era where the quantity of information often outweighs the quality of content. 

Looking ahead, the integration of LLMs in information retrieval systems presents both opportunities and challenges. While LLMs offer a more intuitive and efficient search experience, they also require ongoing considerations regarding accuracy, bias, and the ethical use of AI in information access. The field must continue to evolve, striking a balance between technological innovation and a steadfast commitment to ethical and responsible information dissemination. 

In conclusion, the evolution of information retrieval, driven by the integration of Large Language Models, is not just a technological achievement; it reflects the human pursuit of knowledge and the constant endeavor to refine how we access and engage with the vast realm of information. As this field continues to evolve, it will undoubtedly reshape our relationship with information and the methods we employ to retrieve it.

\bibliographystyle{unsrtnat}
\bibliography{references} 

\end{document}